\documentclass{PoS}

\usepackage{graphicx}  
\usepackage{dcolumn}   
\usepackage{bm}        
\usepackage{amssymb}   
\usepackage{amsmath}
\usepackage{xspace}
\usepackage{epstopdf}
\usepackage{wrapfig}
\usepackage{color}
\usepackage{bbm}
\usepackage{color}

\usepackage{ulem} 
\usepackage{subfigure}

\usepackage{times}
\usepackage{color}
\usepackage{ulem}

\newcommand{\s}{$S$}

\newcommand{\be}{\begin{equation}}
\newcommand{\ee}{\end{equation}}

\title{6-quark Dark Matter}

\ShortTitle{6-quark Dark Matter}

\author{\speaker{Glennys R. Farrar}\\
        Center for Cosmology and Particle Physics, \\
        Department of Physics, New York University, NY, NY 10003 USA
        E-mail: \email{gf25@nyu.edu}}

\abstract{It has recently been proposed that the relatively inert, highly symmetric, neutral flavor singlet scalar hadron made of $uuddss$ quarks may have a mass $< 2 (m_p + m_e)$.  This is consistent with QCD theory, and with existing accelerator and non-accelerator constraints.  For mass in the 1.5-1.8 GeV range, the observed DM relic abundance and the observed DM to ordinary matter ratio can emerge naturally.   Dark matter freezes out before primordial nucleosynthesis and does not significantly impact primordial abundances, so the conventional argument that DM is non-baryonic does not apply.   The interaction cross section between DM and the gas in the Galaxy is such that the dark matter in our local neighborhood is naturally co-rotating with the solar system, to a sufficient degree that DM may not have enough energy to be detected in applicable DM experiments. Interaction with the gas in galactic disks provides the first (non-MONDian) explanation for the striking correlation in the small-scale structure of rotation curves and the inhomogeneous distribution of gas, and also accounts (unlike MOND) for instances of galaxies not exhibiting such correlations.   Depending on the cross-section, a DM-baryon interaction can produce a dark matter disk as suggested by recent studies, and has many or all virtues of self-interacting DM (SIDM) for removing inconsistencies of LCDM.  Lab experiments to discover this particle are discussed.}

\FullConference{35th International Cosmic Ray Conference --- ICRC2017\\
		10--20 July, 2017\\
		Bexco, Busan, Korea}

\begin{document}

\section{A stable sexaquark}
Conventional wisdom, based on the constituent-quark-model for baryon masses, is that the lightest state of matter with baryon number 2 is the deuteron.  However the constituent quark model for states with integral spin has little if any theoretical justification, and is manifestly very wrong for the mesons.   Here we explore the viability and implications of the possibility\cite{fS17} that there is a stable, scalar 6-quark state, $uuddss$.   This state is called sexaquark below, to differentiate it from the the conjectured loosely-bound H-dibaryon \cite{jaffe:H} which has the same quark content but decays in $\approx 10^{-10}$s.  The \s\ can be absolutely stable, with $m_S < 2 (m_p + m_e)$, or effectively stable with a lifetime greater than the age of the Universe, for $m_S < m_\Lambda + m_p + m_e$ depending on the wave function overlap with two baryons\cite{fzNuc03}.  

As elaborated elsewhere \cite{fS17}, isospin/flavor symmetry of QCD means the \s\ cannot exchange pions or other flavor-octets, and consequently it has a much smaller size than octet baryons such as the proton, neutron and $\Lambda$.  One can estimate $r_S \approx 0.15-0.4$ fm, compared to $r_N = 0.9$ fm.   As a result it does not bind to nuclei and limits on exotic isotopes are inapplicable \cite{fzBind03};  constraints on decaying doubly-strange hypernuclei are inapplicable \cite{fzNuc03,ahn+hypernucBNL01,takahashi+hypernucKEK01}; nuclei are stable \cite{fzNuc03}.  Accelerator searches either looked explicitly for decay products (as expected in the H-dibaryon scenario \cite{jaffe:H}) \cite{belz+96,KTeVHdecay99,BelleH13} or were only sensitive to masses $\geq 2$ GeV \cite{NA3longLiveNeuts86,bernstein+LongLivedNeut88}.

Eventually lattice QCD should be able to answer the question of whether the proposed stable sexaquark state exists.  
But while many baryon and meson masses can be accurately calculated, lattice gauge calculations become drastically more difficult with increasing number of constituents (G. P. Lepage, private communication and \cite{LepageTASI}).  Current studies are still very far from the physical limit with respect to quark masses and infinite volume, but the most realistic calculation \cite{Beane+13} using quark masses of 850 MeV/c$^2$, finds 80 MeV binding energy.  This cannot be extrapolated to the few-MeV masses applicable to the relativistic bound state, but it does strongly suggest that a 6-quark state with mass less than $2 m_\Lambda$ should exist.   This lattice QCD evidence pointing to a bound state, combined with the experimental failure to detect a decaying H-dibaryon \cite{belz+96,KTeVHdecay99,BelleH13}, is in fact compelling, albeit indirect, evidence for the stable or effectively stable \s\ as advocated here.

$S$ production can be estimated to be at the level of $\sim 10^{-4}-10^{-6}$ of pion production in high energy hadron collisions such as at the LHC.  Thus many \s's should have been produced in accelerators, but due to their similarity to the much more copiously-produced neutrons, they would go unnoticed in such an environment.  The most effective way to discover such a state is as a peak in missing-mass in $\Upsilon~~[ \rightarrow {\rm gluons}] \rightarrow S \, \bar{\Lambda} \, \bar{\Lambda}~~{\rm or} ~\bar{S} \, \Lambda \, \Lambda$  \cite{fS17}, where the $\Lambda$'s decay and their momenta are reconstructed.  Effectively, resonant production of the $\Upsilon$ in $e^+ e^-$ collisions, provides a copious source of triple-gluon states, which produce quarks, which form hadrons.    These gluons are produced in a region of size $\approx (10 {\rm GeV})^{-1} \approx 0.02$fm and are flavor singlets, giving a relative advantage for producing \s's compared to hadronic collisions.  Furthermore with phase space permitting relatively few accompanying particles, it is much easier to establish the striking $S = \pm 2, B = \mp 2$ apparent strangeness and baryon number non-conservation associated with \s\ production.  Hundreds of millions of $\Upsilon$ decays have been recorded and one can expect that the proposed missing-mass search would be rewarded with discovery, if indeed there is a stable sexaquark which has been missed up to now.

For more details of the particle physics, existing limits, and search strategies, see \cite{fS17}.  The remainder of this presentation will focus on the astrophysics and direct detection of \s\ dark matter, but it should be noted that the discussion is more general and can be applied to any hadronically interacting dark matter candidate.   A discussion of the DM relic abundance and the dark matter to baryon ratio in the SDM (\s\ dark matter) scenario will be presented elsewhere.

\section{Dark Matter detection}
From the estimated $r_S \approx 0.15-0.4\,$fm radius of the \s, we can naively estimate the cross section for an \s\ scattering from nucleons or another \s: $\sigma_{Sp} = \sigma_{Sn} \approx (0.25-1) \sigma_{NN}$, and $\sigma_{SS} \approx (0.25-1) \sigma_{SN}$.  However the true cross sections can be much smaller or larger than suggested by such naive estimates.  Cross sections of known hadrons are roughly geometrical when $v \approx c$, but at low energy they are governed by potential scattering, for which interference and resonance effects can cause major deviations from simple geometric behavior.  E.g.,  the nucleon-nucleon elastic cross section is $\approx 20$ mb for $v \approx c$ but grows to 1500 mb for $v \rightarrow 0$.  

Deep underground detectors provide WIMP direct detection limits far below the hadronic level,  $\sigma_{\chi N} \lesssim 10^{-38} {\rm cm}^2 = 10^{-11}$ mb, for DM masses ${\mathcal O}$(GeV), and lower still for higher DM mass.   However these detectors are insensitive to DM with too-large cross section, due to energy loss en route to the detector \cite{SGED90,zf:windowDM}.  The most sensitive underground experiment for cross-sections in this range is DAMIC \cite{DAMIC12} operating 107 m below the surface.  At this depth, scattering in the Earth's crust and the detector shielding reduces the flux of detectable DM particles reaching the DAMIC detector by more than a factor $\sim 10^8$ for a $\sim$ GeV DM particle with $\sigma_{\chi N}  \geq 0.7 \times 10^{-29} {\rm cm}^2$ \cite{mfBounds17}.  (Note that the previous analysis of DAMIC by \cite{kouvarisShoe14} overestimated the attenuation by more than a factor of $10^4$ and thus underestimated the sensitivity range of DAMIC \cite{mfBounds17}.)

To access higher cross sections requires minimal overburden, and the best detector is the X-ray Quantum Calorimeter (XQC) \cite{xqc}, aboard a sounding rocket in the upper atmosphere.  Successively more detailed limits using XQC data have been provided by \cite{wandelt:SIDM, zf:windowDM,erickcek+07,mfBounds17}.  
As shown in \cite{mfBounds17}, the correct limit on the cross section from XQC, assuming the standard DM velocity distribution and taking the XQC sensitivity at face value, is $\sigma_{\chi p} \leq 0.6 \times 10^{-30} \, {\rm cm}^2$.  This is a factor 9 more stringent than found in \cite{erickcek+07}, due to  \cite{erickcek+07} mistakenly assuming the rocket body shielded the detector.   With the poor approximations in \cite{kouvarisShoe14} and \cite{erickcek+07} corrected, the window that previously existed for DM to have $\sim \mu$b cross section with nucleons, is closed \cite{mfBounds17}, with the above-mentioned caveats.  The window was in any case at too-low cross section values to be relevant for the \s, for which we expect $\sigma_{SN} \geq 5$ mb.   

From simple kinematics, the fractional energy deposit when a DM particle of mass $M_{\chi}$ scatters off a target nucleus of mass $m_t$ at rest is
\be
\label{Edep}
 f = \frac{2 \, m_t \, M_{\chi}}{(m_t + M_{\chi})^2} \,(1 - {\rm cos}\, \theta),
\ee
where $ \theta$ is the center-of-mass scattering angle.  Equating the maximum energy deposit in the Si-based quantum calorimeter (XQC) to their 29 eV threshold, gives the minimum relative DM velocity for XQC to have any sensitivity, which for DM mass $ M_{\chi}<< 28 m_p$ is:
\be
\label{vmin}
v_{\rm min, XQC} =  99 \,{\rm km/s} \left( \frac{2 \,m_p}{M_{\chi}} \right) .
\ee
The lower the mass of the DM, the larger its velocity must be relative to the detector to deposit the required energy.  A typical assumption for the DM velocity distribution in the Galactic rest frame, is an isotropic Maxwellian with peak at $220$ km/s and cutoff at the escape velocity 584 km/s \cite{erickcek+07,mfBounds17}, resulting in a DM velocity distribution with respect to the Earth which peaks at $v_p \approx$300 km/s.  
For low DM mass, the limit is very sensitive to $v_p$.  E.g., reducing $v_p$ to $\approx 40$ km/s would be sufficient to weaken the XQC bounds to $\sigma_{\chi N} \approx 0.1\,$b for $M = 2 \, m_p$ \cite{mfBounds17}. 

\section{Co-Rotation of Dark Matter}
The typical velocity dispersion of local stars and gas relative to the sun is 10-20 km/s~\cite{dehnenBinney98,Lopez-Santiago06,KalberlaKerp09}.   Thus if DM interactions were sufficient to bring local DM into co-rotation at a level comparable to local baryons, the DM energy deposits in XQC would be far below threshold.  

Taking $M = 2 \, m_p$ as a benchmark, Eq. (\ref{Edep}) shows that the average fractional energy loss per DM collision with H or He is 4/9.  
 
\begin{figure}[t]
	\centering
	\includegraphics[trim = 0.in 0.0in 0.in 0.in, clip, width=0.48\textwidth]{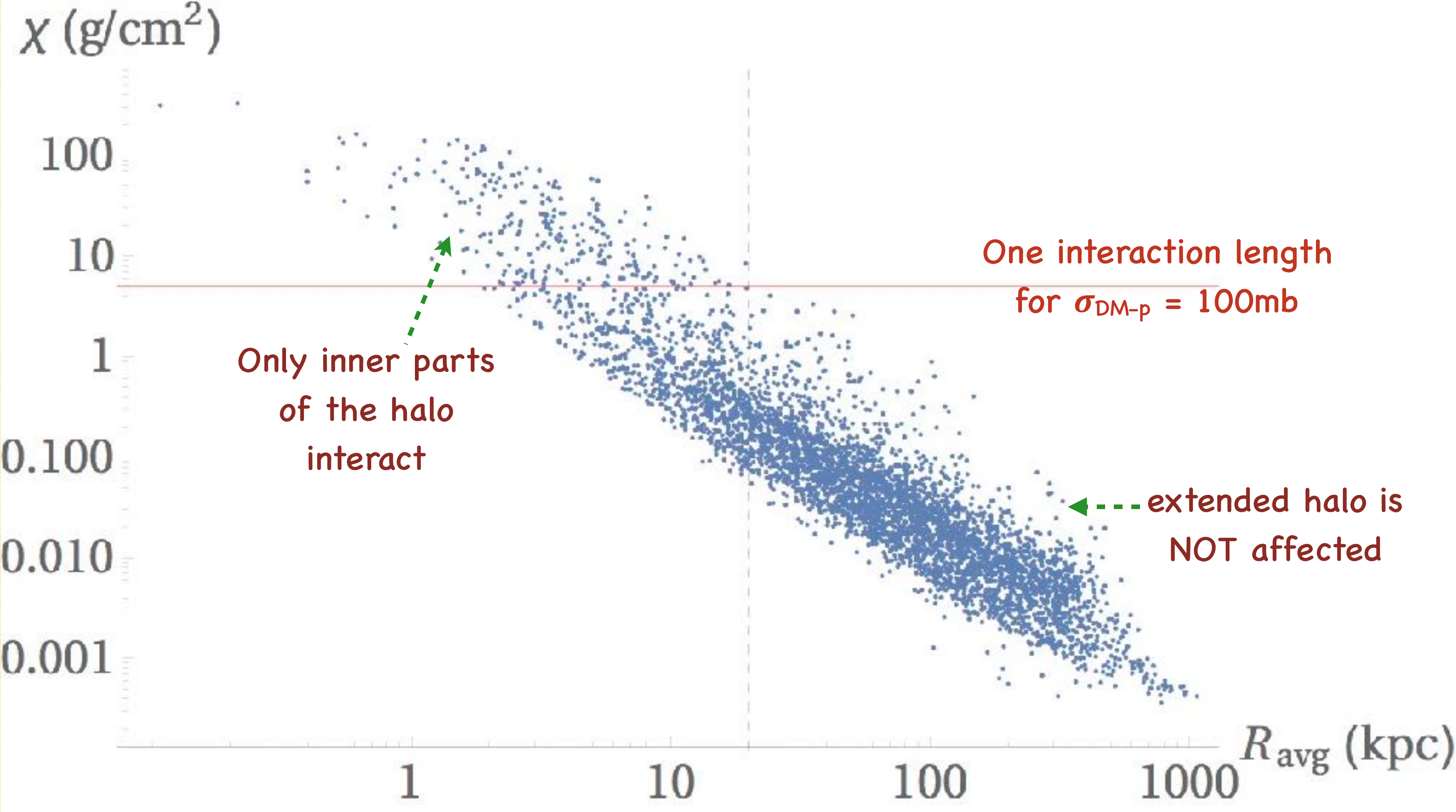}\label{colDepth}
	\caption{ Column Depth accumulated in 10 Gyr vs average radius of orbit, in units of g/cm$^2$, figure courtesy D. Wadekar.}
  \end{figure}
 
D. Wadekar and I have been carrying out simple simulations to understand if DM may come into sufficient degree of co-rotation to significantly impact direct detection limits, and if so for what range of $\sigma_{\chi N}$.   The first step in assessing plausibility is to estimate the number of interactions which would be experienced by DM particles in different regions of the halo.   Using the code SMILE \cite{SMILE}, which implements the Schwarzschild method, we generate an ensemble of DM orbits self-consistently producing a stable DM halo for the standard choice of NSF+baryonic disk Milky Way potential provided in \textit{galpy} \cite{BovyGalpy15}.   Propagating these orbits through the Galaxy, we record the accumulated gas column depth as a function of time.  Fig.  \ref{colDepth} shows the accumulated column-depth after 10 Gyr, $X = \int \rho d \ell$, as a function of the average radius $R_{\rm avg}$ of the (non-interacting) orbit, using the standard baryon distribution in \textit{galpy}.  (Note that when DM particles interact, their orbits change -- they are typically drawn closer to the disk and inward -- so the column depth distribution for true, interacting orbits in the inner region will be larger than in Fig. \ref{colDepth}.)  Orbits with large $R_{\rm avg}$ accumulate little $X$ and are thus unlikely to interact because they spend most of their time at large radii, where there is little gas, while orbits living predominantly in the region of the solar radius see a much larger column depth of gas.   

Because DM at large average radius has a low interaction probability, the large-scale structure of the halo is unaffected by hadronic-level DM-nucleon interactions and will be unchanged with respect to LCDM modeling; thus it agrees with observations.  On the other hand, orbits with $R_{\rm avg} \leq 10$ kpc have sufficient interactions even in our simplified study to become co-rotating for $\sigma_{\chi p} \gtrsim 200$ mb\footnote{The probability of colliding with a He nucleus is 5.4 times larger than with a p, because the cross section scales as $(A \times {\rm reduced \,mass})^2$ and $n_{\rm He} = 0.078 \,n_{\rm H}$.  As noted earlier, $\sigma_{NN} \approx 1500$ mb in this velocity range and thus $\sigma_{\chi N} \gtrsim 200$ mb may be reasonable.}.   The actual cross section required for co-rotation may be less, because the column depth is higher with detailed, observation-based gas models and because the Galaxy grows through accretion so modeling it a static DM and gas distribution is not accurate.    Additionally, ref. \cite{read+DD08} argues that as a result of merging sub-halos being stochastically dragged into the disk, there could be a thick disk of DM, with local mass density 0.2-1 times the local density for a conventional DM halo.  This would be expected to co-rotate to some extent, perhaps with an estimated lagging circular velocity $\sim 50\,$km/s and a velocity dispersion of $\approx 50$ km/s.  Hadronically interacting DM particles originating in such a thick dark disk would have a head-start in approaching the gas-disk phase space distribution.

Ref. \cite{wfCoRotation17} reports results of more detailed studies following the trajectories of DM particles -- initially on SMILE non-interacting DM orbits -- through the Galaxy, randomly generating the next interaction point, where they scatter isotropically in the center-of-mass off a H or He nucleus which is rotating with the gas rotation curve.  (Stars have such a small filling factor that they do not contribute significantly to causing the DM to co-rotate, so we only consider interactions with the gas.)   We find that for cross sections as estimated above, most of the DM at the solar radius comes into sufficient co-rotation to escape detection while at large radii the DM distribution is unaffected.  The impact of time evolution and spatial structure in the gas, and DM self-interactions, will also be reported in \cite{wfCoRotation17}.  

\section{Astrophysical Hints for a DM-baryon interaction}
    
As a result of the DM-gas interactions, the DM and gas phase space distributions approach one another.  Baryonic processes dominate the gas distribution, so the mild heating of the baryons in the Milky Way disk due to DM interactions will be difficult to verify in our own Galaxy until baryonic physics is much better controlled than at present.   But the heating predicted in this model -- roughly comparable to that due to SNe explosions -- is welcome for explaining the quenching of star formation in galaxies in general.  This is an outstanding problem in cosmology for which AGN feedback and the known SN rate seem insufficient, so an additional source of heating is helpful.  

Several papers in recent years have argued for the existence of a disk of Dark Matter, embedded within the standard roughly spherical DM halo.  The most recent paper is based on paleoclimatic evidence \cite{shavivPaleo16};  see \cite{KRandallgasDarkDisk16} for a list of other references.   A thin DM disk can form if a portion of the DM is capable of cooling, e.g., as in \cite{fan+Randall+DDDM13}.  A thick DM disk is argued to be a natural consequence of the Galaxy's growth through accretion \cite{read+DD08}.  A DM-nucleon interaction also produces a disk, whose thickness depends on  $\sigma_{\chi N}$ and $M_\chi$;  the relationship is explored in \cite{wfCoRotation17}. 

Another effect of a DM-gas interaction is the development of structure in the dark matter of the inner Galaxy, reflecting the structure in the gas.   Ubiquitously, across all types of disk galaxies, the slope of the inner rotation curve follows the baryonic matter, simply amplified on account of having non-negligible contribution from DM.  Thus the DM distribution follows the baryonic one, rather than being universal independent of the baryons as expected from standard LCDM\cite{lelli+13}. 

An even more specific hint for DM-nucleon interaction comes from two different types of observed correlations between features of rotation curves, and the distribution of baryons:\\  
 $\bullet$   There is a close relationship between baryons and the rotation curves in the inner region of the galaxy, e.g., \cite{PalunasWilliamsMaxDisk00}; see \cite{swaters+12} for references to more recent literature.  This is not simply the core-cusp problem, which requires the DM density in the inner region of the galaxies to be smoother than the NFW function.  Self-interacting DM can account for smoothing \cite{ss:SIDM}, but does not produce a correlation between the slope of the inner rotation curve and that of the baryon distribution.
 \\
 $\bullet$ Detailed rotation curve measurements show that many galaxies' rotation curves exhibit wiggles which match the shape of wiggles due to the gravitational potential of the observed stars and gas, over a wide range of galaxy types \cite{swaters+12}.   
The effect is illustrated in Fig. 
2.  The first two panels are just a few from many galaxies reported in  \cite{swaters+12},  showing rotation curves of the galaxies along with the baryonic contribution, rescaled to give the best fit the rotation curve.   In some cases the match between the shapes is astonishingly detailed.  There is some scatter in the required rescaling but it is relatively narrow within a given galaxy type \cite{swaters+12}.  This shape correlation has been cited as evidence for MOND.  But cases like UGC7399 and other examples not shown such as UGC8490, where structure in the gas is \emph{not} reflected in the rotation curves, invalidate that interpretation because in MOND there should be no exceptions to the correlation.   The last panel shows an example from different authors \cite{creasey+17} which manifests the same phenomenon, namely that the fit would be better if the DM contribution reflected the shape of the gas contribution.
 
 If there is a DM-baryon interaction as contemplated here, then after sufficiently many scatterings the DM would be expected not only to co-rotate and form a disk, but to form density substructure resembling that of the gas --  because locally the DM orbits converge to the orbits of the gas particles they scatter from.   However when galaxies are disrupted by a major merger, the DM distribution gets ``reset" to be smooth.  Spiral arm structures in the gas arise soon after the disk reforms, whereas many orbits are required for the DM to follow.  This may naturally account for the presence of shape correlations in some galaxies and its absence in others.  It predicts that galaxies with non-matching rotation curves may show evidence of a recent major merger.  This is in fact the case for UGC7399 and UGC8490, as is visible from their images.  
 

 
\begin{figure}[t]
	\centering
\includegraphics[trim = 0.1in 4.8in 0.2in 0.3in, clip, width=\textwidth]{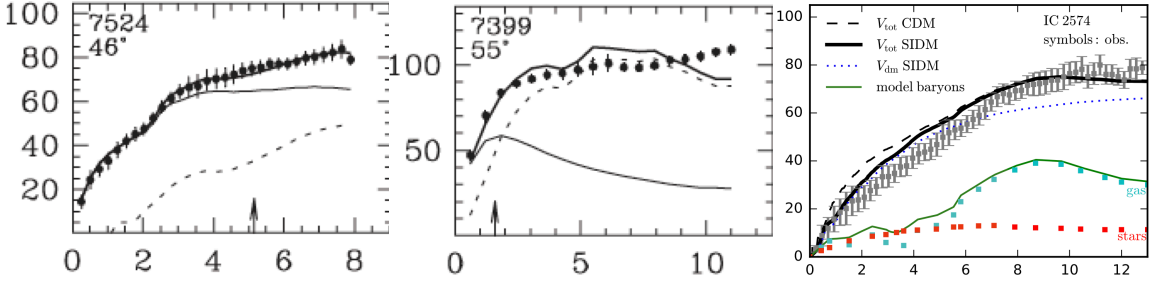}
	\vspace{-0.35in}
	\caption{ Rotation curves from (left \& center) Fig. 1 of ref. \cite{swaters+12}  and (right) Fig. 6 of ref. \cite{creasey+17}.  Gas densities are shown as dashed lines \cite{swaters+12} and green squares \cite{creasey+17}.  UGC7524 (left) and IC2574 (right) are examples of rotation curves with a distinctive non-smooth structure suggestive of a scaled-up version of the gas density distribution, while UGC7388 (center) has a smooth rotation curve in spite of the bumps in the gas density profile.}\label{RotCurveEx}
	\vspace{-0.15in}
  \end{figure}
  

\section{Acknowledgements}  This research was supported by NSF-PHY-1212538 and the James Simons Foundation.  The co-rotating dark matter scenario is being studied in collaboration with D. Wadekar; some results of that work are presented here.   Further work refining XQC limits in this scenario is underway with M. S. Mahdawi.   Many other colleagues have generously provided information and suggestions. 

\def\apj{Astrophys.\ J.}
\def\nat{Nature}
\def\apjl{Astrophys.\ J. Lett.}
\def\apjs{Astrophys.\ J.\ Supp.}
\def\aap{Astron.\ Astrophys.}
\def\prd{Phys. Rev. D}
\def\physrep{Phys.\ Rep.}
\def\mnras{Month. Not. RAS }
\def\araa{Annual Rev. Astron. \& Astrophys.}
\def\aapr{Astron. \& Astrophys. Rev.}
\def\aj{Astronom. J.}
\def\jcap{JCAP}

\providecommand{\href}[2]{#2}\begingroup\raggedright\endgroup

\end{document}